\documentclass[12pt,preprint]{aastex}

\usepackage{natbib}

\slugcomment{ApJ, in press \today}

\begin{document}

\title{The density of mid-sized Kuiper belt object 2002 UX25 and the 
formation of the dwarf planets}
\author{M.E. Brown}
\affil{Division of Geological and Planetary Sciences, California Institute
of Technology, Pasadena, CA 91125}
\email{mbrown@caltech.edu}

\begin{abstract}
The formation of the largest objects in the Kuiper belt, with
measured densities of $\sim$1.5~g~cm$^{-3}$ and higher, from the coagulation
of small bodies, with measured densities below 1~g~cm$^{-3}$
is difficult to explain without invoking significant porosity in
the smallest objects. If such porosity does occur,
measured densities should
begin to increase at the size at which significant porosity is no longer 
supported.
Among the asteroids, this transition occurs for 
diameters larger than $\sim$350 km. In the Kuiper belt, no density 
measurements have been made between $\sim$350 km and
$\sim$850 km, the diameter range where porosities might 
first begin to drop.
Objects in this range could provide
key tests of the rock fraction of small Kuiper belt objects.
Here we report the orbital characterization, mass, and density determination of
the 2002 UX25 system in the Kuiper belt. For this object, with a diameter of $\sim$650 km, we find
a density of $0.82\pm0.11$ g cm$^{-3}$, making it the largest solid
known object in the solar system
with a measured density below that of pure water ice.  
We argue that the porosity of this object is unlikely to be
above $\sim$20\%, suggesting a low rock fraction.
If the currently measured densities of Kuiper belt 
objects are a fair representation of the sample as a whole, 
creating
$\sim$1000 km and larger Kuiper belt objects with rock mass fractions
of 70\% and higher from coagulation of small objects with rock fractions
as low as those inferred from 2002 UX25 is difficult.
\end{abstract}

\keywords{Kuiper belt: general --- Kuiper belt objects: individual (2002 UX25) --- planets and satellites: formation}

\section{Introduction}

In standard accretionary scenarios for growth of objects
in the Kuiper belt \citep[i.e.][]{2008ssbn.book..293K}, the objects all
form in
regions of the nebula with similar physical characteristics
and so should be composed of roughly similar amounts of rock and ice.
It was surprising, therefore, to find that measured densities
of Kuiper belt objects (KBOs) range from as low as 0.5 g cm$^{-3}$ 
to at least 2.6 g cm$^{-3}$ \citep{2012AREPS..40..467B}.

A clear trend has emerged: the smaller objects have low densities, while 
larger objects have increasingly higher densities \citep{2006ApJ...643..556S,
2007Icar..191..286G,2012A&A...541A..94V,2013A&A...555A..15F}.
While larger objects often
have higher densities due to pressure-induced phase changes, ice-rock bodies
like these need to approach the size of Triton before this effect becomes
significant  \citep{1979Icar...40..157L}. A more likely culprit for the
low densities of the small objects is porosity. Nothing is known about the 
porosity of KBOs, but in the asteroid belt the average
porosity for objects as large as $\sim$350 km in diameter -- the size of the
small KBOs -- has been calculated to be $\sim$50\% \citep{2011AJ....141..143B}, 
meaning that the compressed density of the object
would be a factor of two higher than the measured density. Ice and rock 
compression experiments show that rock is capable of supporting much
more porosity \citep{2009JGRE..114.9004Y}, so we regard asteroid porosities as a plausible upper
limit to the porosities of icy KBOs. For this maximum porosity,
the weighted average density for small KBOs of 0.6 g cm$^{-3}$ corresponds to 
porosity-free density of $\sim$ 1.2 g cm$^{-3}$. 
Even for a high assumed porosity, 
these small KBOs have only about one third rock by mass. 
Simple coagulation of these rock-deficient 
objects will not lead to the much higher densities of the rock-rich dwarf 
planets.

With porosity the key unknown for small KBOs,
an important clue to the formation of the dense 
dwarf planets would be the measurement of the densities of the
smallest KBOs which are large enough to have had most
of their porosity compressed out.
Here we report the detection and orbital characterization of a satellite
to the $\sim$650 km
diameter hot classical KBO 2002 UX25. This object is an order of magnitude
more massive than the next largest KBO with a measured density.
We consider the effects of porosity
on 2002 UX25, and we use the density we calculate for
this system 
to attempt to understand the formation mechanism of the dwarf planets.

\section{The orbit of the satellite of 2002 UX25}
\subsection{Hubble Space Telescope observations}
A satellite of 2002 UX25 was discovered in observations from the
High Resolution Camera (HRC) of the Advanced Camera for Surveys (ACS)
of the Hubble Space Telescope (HST) on 26 August 2005.
In order to determine the orbit of the satellite 
and the mass of the system, we obtained
a series of 6 follow up observations a year later. For each
observation we obtained
8 exposures of 275 second duration using the F606W filter.

\begin{figure}
\plotone{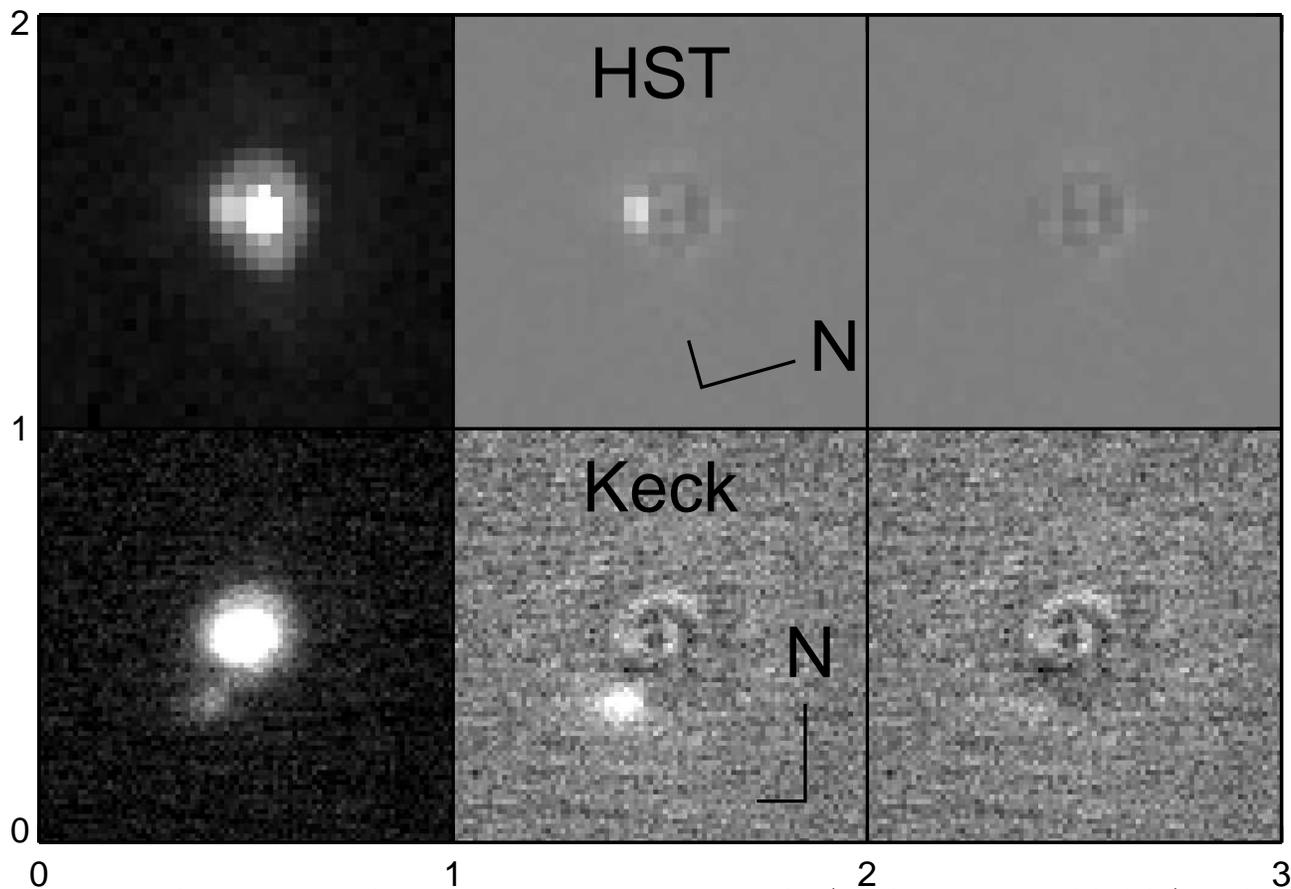}
\caption{Observations of the 2002 UX25 system with HST/HRC 
and Keck LGS-AO/NIRC2. The northward orientation arrow is 0.25 arcseconds 
long, for scale. In the first column, we show the image of both 2002 UX25 and its satellite.
From this image we simultaneously fit a PSF to both the primary and satellite. In the second column
we show the image with the primary part of the fit subtracted. In the final column we show both 
components subtracted. The HST observation is from JD 2453939.98322 and is the most
blended of the detections.
}
\end{figure} 
The satellite is cleanly detected in 4 observations, undetected in
2 observations, and detected but blended with the primary
in 2 observations (Figure 1).
Astrometric positions of the 
satellite relative to
2002 UX25 were obtained following the method of \citep{2010AJ....139.2700B}, 
in which a five-times oversampled theoretical point spread function (PSF)
is constructed for the pixel location of 2002 UX25 
using 
TinyTim \citep{Krist1993}, the HST PSF modeling software, and then the
sub-pixel centers of 2002 UX25 and the satellite, 
the total flux of 2002 UX25 and the satellite, and the
sky background are optimized using an iterative least-squares fit. 
We determine the uncertainties for each observation
from the scatter of the positions measured in the 8 individual images
acquired within a single HST orbit. 
We often detect motion consistent with the satellite orbital
velocity within single sets of observations.
To be conservative, however,
we assume that all deviation within one orbit is due only to
measurement error. Even in the most blended observation, we obtain consistent 
measurements in all 8 of the individual images during a single visit.
The astrometric positions of the satellite are given in Table 1.

\subsection{Keck laser guide star adaptive optics observations}
The HST astrometric observations 
lead to a mirror ambiguity in the determination of the 
orbit pole (see below). For the 2002 UX25
system, breaking this ambiguity is particularly important; one orbital configuration would be undergoing current
mutual events, while the other had its mutual event season before
the satellite discovery. A single well placed astrometric point several years later 
could break this ambiguity.
We obtained a single astrometric point using laser guide star adaptive optics (LGS AO)
at the Keck Observatory \citep{2006PASP..118..297W, 2006PASP..118..310V} on 7 December 2012. Observations were 
scheduled for a night when 2002 UX25 passed within 35 arcseconds of an $R\sim13.7$ 
star that could be used for tip-tilt correction. We obtained a total
of 73 individual 2 minute integrations of the system using the NIRC2 camera 
with a 0.02 arcsecond plate scale and the K$_p$ filter. Image full-width-half-max (FWHM) measured
on 2002 UX25 ranged from 70 to 110 mas, worse than the 45 mas theoretical
diffraction limit of a 10-meter telescope, but consistent with typical
LGS AO performance with a moderate brightness off-axis tip-tilt star.

The satellite was visible in the best single 2 minute exposures and easily visible in 
all medianed stacks of five exposures 
(Figure 1). 
To accurately determine the astrometric position
of the satellite we first selected the images with the 
best LGS AO correction. We determined the quality of the correction
 by fitting a single two-dimensional gaussian function at the position
of the primary and calculating the average FWHM of the core. We then retained only
the half of the data with a correction above the median value. 
These data were shifted to
place the primary at a common position and then median-combined 
into 6 groups of 5. 

While the satellite is outside of the core of the PSF,
it is
within the halo, which could affect measurements of its position. 
To accurately measure the position,
we fit the primary and satellite
with a PSF model that is the sum of two arbitrarily oriented two-dimensional Gaussian distributions.
The residuals from these fits at the location of the satellite are nearly indistinguishable from 
background noise, thus the astrometric fits to the satellite position
will no longer be affected
by the halo of the primary PSF.
As with the HST data, we determine the errors 
in the astrometric and photometric fits from the dispersion of the measures
in the individual stacked frames (Table 1). 

\subsection{Orbit fits}
\begin{figure}
\plotone{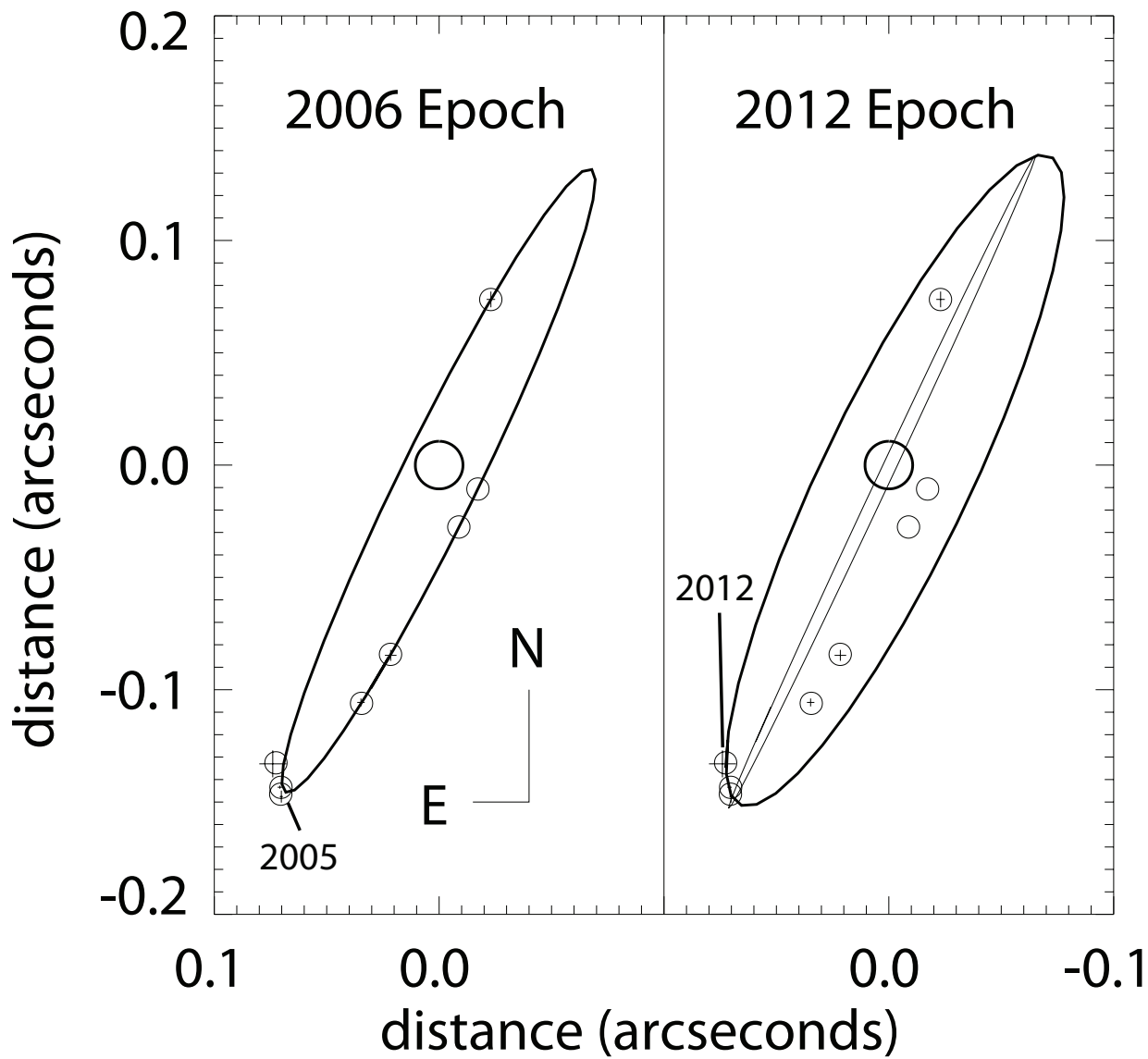}
\caption{The orbit of the satellite of 2002 UX25 at the epoch of discovery
and at the current epoch. The small circles show the predicted location of
the satellite based on the best fit of all data from 2005-2012, while the crosses 
inside these circles
show the observations and their uncertainties. For better visibility, 
the 3$\sigma$
uncertainties are shown. For the 2012 epoch, we
show both the best fit orbit (thick line) and the mirror image orbit (thin line)
which we exclude with high confidence. The large circle in the center shows the
approximate size of 2002 UX25, while the small circles show the size of the satellite.
}
\end{figure} 
It appears that the 
2002 UX25 satellite is close 
to being in an edge-on orbit (Figure 2). Such an orbit is consistent with the 2
 non-detections of the satellite in the HST data. 
We 
determine the best-fit orbit to the observations by using a 
Powell scheme to minimize the 
$chi^2$ value of the residuals and find the optimal orbital 
parameters. We ignore the non-detections,
and note that all good fits naturally place the non-detections
too close to the primary to have been observed. 
Our elliptical orbit has free parameters
of semimajor axis, 
orbital period, inclination, longitude of the ascending node, mean anomaly,
longitude of perihelion, and angle of nodes.
The best fit has
a $\chi^2$ value of 2.97 , or a reduced $\chi^2$ for 5 degrees of freedom 
(6 sets of $x$, $y$ coordinates
minus 7 orbital parameters) of 0.6, suggesting that the uncertainties have
indeed been overestimated. Forcing a fit to the mirror image orbit, 
we find a $\chi^2$ value of 12.0, nearly four times higher than the best fit. 
As expected, the 2012 astrometry is the main discriminant between the two orbits.
While the best fit orbit fits a position within 15 mas of the 2012 astrometric point, 
the mirror image orbit deviates by 92 mas, well outside of the uncertainty of the observation
(Fig 2).
We conclude
that we have resolved the mirror ambiguity and found the true
orbital solution. The satellite plane crossed the line of sight to the earth and had mutual
events in 2001. The next mutual event season does not occur until 2109.

The best-fit orbit has a moderate eccentricity of 0.17. Attempting a circular fit gives
a best-fit $\chi^2$ of 56.5, or a reduced $\chi^2$ for 7 degrees of freedom of 8.1, significantly
higher than for the elliptical fit. We conclude that the orbit is indeed elliptical.

We explore the uncertainties on the eccentricity and the other parameters 
by integrating through phase space using a 
Markov Chain Monte Carlo (MCMC) scheme. We use the Python 
package {\it emcee} \citep{2012arXiv1202.3665F} which implements the 
\citet{ISI:000282653600004}  affine invariant ensemble sampler for MCMC. We assign 
uniform priors on all parameters (with two parameters being $e\sin\omega$ 
and $e\cos\omega$, where $e$ is the eccentricity and $\omega$ is the argument 
of perihelion, rather than simply $e$ and $\omega$) 
and find good convergence with an
ensemble of 100 chains running 10$^4$ steps with a initialization
(``burn-in'') period, which is discarded,
 of 10\% of the total length of each chain. The marginalized distribution
of each of the parameters is nearly gaussian, we thus report the median and the middle
68.2\% to represent the best fit plus 1$\sigma$ uncertainties (Table 2). 
We derive a marginalized
system mass of $1.25 \pm 0.03 \times 10^{20}$ kg, or about 0.7\% the mass of Eris, the most massive known 
object in the Kuiper belt.

\section{The size, density, and tidal evolution of 2002 UX25}

Thermal radiometry of 2002 UX25 and its satellite has been obtained 
from both the Spitzer and Herschel Space Telescopes \citep{2008ssbn.book..161S,2013A&A...555A..15F}.
A combined
analysis using measurements at wavelengths from 24 to 500 $\mu$m  
suggests an effective
diameter of the system of 692$\pm$23 km with an albedo of 10.7$^{+0.5}_{-0.8}$\%. 

As the fractional brightness of the satellite measured with HST 
at visible wavelengths is comparable to
that measured with Keck AO at infrared wavelengths (Table 2),
with a mean value of 8\% , it is plausible that 2002 UX25 and its satellite share the
same surface characteristics and thus the same albedo. In this case 
we can estimate their separate diameters by assuming that the total thermal emission from each
is proportional to their surface areas. Thus, the diameter of the primary would be
$\sim$664 km while the satellite would be $\sim$190 km. Alternatively, if the
satellite is assumed to have a $\sim$5\% albedo typical of some smaller non-cold classical KBOs \citep{
2012A&A...541A..93M,2012A&A...541A..94V,2012A&A...541A..92S}, the sizes
would be $\sim$640 and $\sim$260 km, respectively. Assuming that the densities
of the primary and satellite are identical, the densities for these two cases would be
0.79$\pm$0.08 and 0.85$\pm$0.08 g cm$^{-3}$, respectively.
For simplicity, we will report the single average value with the
full uncertainty range as 0.82$\pm$0.11 g cm$^{-3}$.
2002 UX25 is the largest object in the Kuiper belt 
with a measured density lower than 1 g cm$^{-3}$.

The eccentricity of 0.17$\pm$0.03 appears unusual for such a large object
with a close satellite.
For the size of 2002 UX25 and its satellite, we can estimate a time scale for
damping of eccentricity by tides of the satellite of 
$$\tau = -{e\over{\dot e}}= {4\over 63}\ {m_s\over m_p}\ ({a\over r_s})^5\ {\mu_s Q_s \over n},$$
where $m_s$ and $m_p$ are the satellite and primary masses, $a$ is the semimajor axis, $r_s$ is the
radius of the satellite, $Q$ is the tidal
quality factor, $n$ is the orbital angular frequency, and $\mu_s$ is the effective rigidity 
of the satellite, defined as
$$\mu_s= {{19 \mu}\over{2p\ g\ r_s}},$$
where $\mu$ is the material rigidity, $\rho$ is the satellite density, and $g$ is the satellite
surface gravity  \citep{MandD}. 
We find for 2002 UX25 an eccentricity damping time scale of $\sim 4 ({\mu\over 4\times 10^9 {\rm N~m}^{-2}})({Q\over100})$ Gyr,
comparable to the age of the solar system for these reasonably assumed values of $\mu$ and $Q$ \citep{MandD}.
The moderate eccentricity of the satellite of 2002 UX25, then, appears a reasonable outcome if
the formation mechanism yielded an initially eccentric orbit or if eccentricity excitation ever occurred in the
past.

\section{The densities of the Kuiper belt objects}
\begin{figure}
\plotone{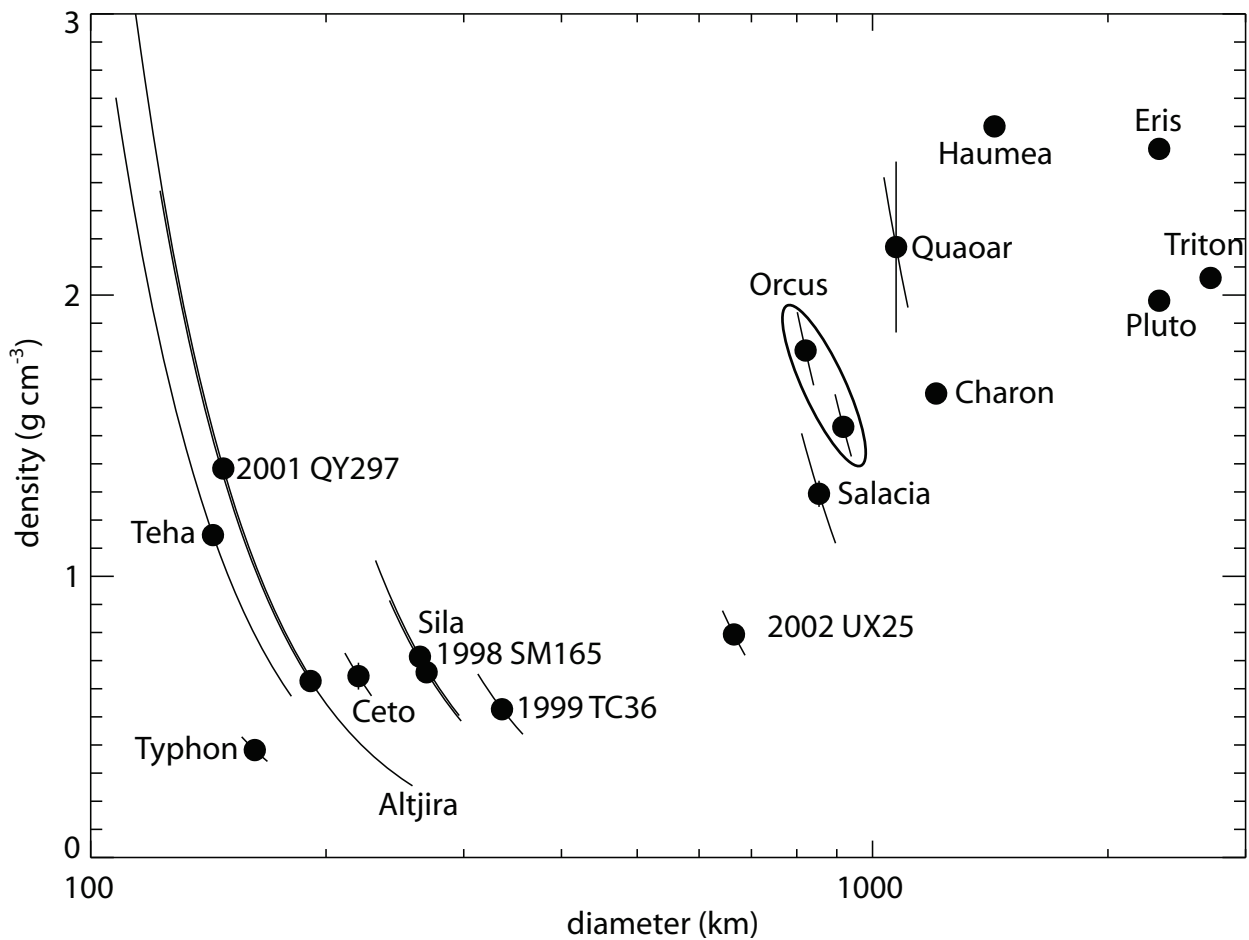}
\caption{Densities of objects in and from the Kuiper belt. In most cases,
the uncertainty in diameter is much larger than the uncertainty in mass,
so the density-diameter uncertainty lies along a curved path. Quaoar has 
a larger mass uncertainty than most other objects, and that full uncertainty
is show as a vertical error bar at the position of Quaoar. Two possible
density-radius solutions are show for Orcus, one where Orcus and its
satellite Vanth have equal albedos (the less dense solution)
and one where Vanth has a lower 
albedo more typical of smaller KBOs (the more dense solution).}
\end{figure} 
We construct the
size-density relationship for all objects with measured masses and sizes (Figure 3).
We assume equal albedos and densities for all of the bodies in the system 
and derive the diameter and density of the primary object from the system mass 
and the measured effective diameter. System masses are taken from \citet{2006ApJ...639.1238R}, \citet{2006AJ....132..290B}, 
\citet{2007Sci...316.1585B}, \citet{2007Icar..191..286G}, \citet{2008Icar..197..260G}, 
\citet{2010Icar..207..978B}, \citet{2010AJ....139.2700B}, 
\citet{2011Icar..213..678G}, \citet{2012Icar..220...74G}, \citet{2012Icar..219..676S} and \citet{2013Icar..222..357F},
while effective diameters are from \citet{2008ssbn.book..161S}, \citet{2012A&A...541A..94V}, \citet{2012A&A...541A..93M}, \citet{2012A&A...541A..92S}, 
and \citet{2013A&A...555A..15F}, with the combined Spitzer-Herschel results being used whenever available.
For Orcus we derive densities
for both the case where the albedos of the primary and satellite are assumed to be 23\% \citep{2013A&A...555A..15F}
and for the case where the satellite has a more typical lower albedo of
5\%, leading to a higher density for the system. 

The low density of 2002 UX25
places strong constraints on any hypothesis proposed for
the cause of the KBO size-density relationship. 
Objects of this
size in the asteroid belt have porosities of $\sim$20\% and lower
\citep{2011AJ....141..143B}
In the Kuiper belt, porosities of objects this size should be lower; ice is 
more compressible at higher pressure \citep{2009JGRE..114.9004Y}, and {\it much} more compressible
if the internal temperatures are elevated. Models of the internal structure
of KBOs of this size range usually conclude that enough internal heating
has occurred from radioactivity and accretional heating that liquid water
is present at some point in the history of the object
\citep[see review by][]{2008SSRv..138..147P}. Bulk porosities, in 
that case, will be low. 

While true porosities of cold icy large objects remain unmeasured, the
analogies to stronger asteroids, the laboratory experiments, and the 
internal modeling all
suggest that 2002 UX25 should not support significant bulk porosity.
Assuming an upper limit of 20\% for the porosity 
gives a compressed density of
2002 UX25 of close to 1 g cm$^{-3}$. Unless we have severely underestimated
the porosity for this object, the rock fraction of 2002 UX25 is similar to
the low rock fraction of the smaller KBOs.

\section{Conclusions}
The inferred low rock fraction of the 2002 UX25 system makes the formation
of rock rich larger objects difficult to
explain in any standard coagulation scenario. For example, to create an
object with the volume of Eris would require assembling $\sim$40 objects
of the size of 2002 UX25. Yet the assembled object, even with the additional
compression, would still have a density close to 1 g cm$^{-3}$ rather than
the 2.5 g cm$^{-3}$ density of Eris \citep{2011Natur.478..493S}.

We offer a small number of possible ways in which the dwarf planets could
still be created from the coagulation of smaller KBOs. First, it is 
possible that we have severely underestimated porosities. If 2002 UX25
could support a porosity of 50\%, it would have a compressed density 
similar to that of Orcus or Charon. If the smaller objects have porosities
of 60\% or higher, they too would have a similar rock fraction to the smaller
dwarf planets and coagulation would no longer present difficulties. 
The inferred change porosity from 2002 UX25 to Salacia to Orcus, over a
relatively small range in diameter, would be unexpected. While such an
extreme porosity for 2002 UX25 cannot be excluded, asteroid observations,
internal modeling, and laboratory compression experiments all suggest that
this possibility is unlikely.

The second manner in which dwarf planets could be built from small
bodies is if the objects for which we have measured densities are
not a fair sample of the Kuiper belt. 
Many -- but not all -- of the low density objects are
part of the cold classical Kuiper belt, which is known to have many
distinct physical properties, including a larger fraction of satellites
and thus a tendency to have a measured density \citep{2008Icar..194..758N}. 
The objects 
Typhon and Ceto are both
Centaurs, however, which are unlikely to be escapees from the stable
cold classical region. The object 1998 SM165 is currently in a 2:1 resonance
with Neptune, so its initial origin is more ambiguous. 2002 UX25, however, with
an inclination of 19 degrees appears to be a clear member of the hot classical
population. It is thus
clear that low density
small objects exist in the non-cold classical population of the Kuiper belt.

Another possibility is that there is a bias in our density measurements.
If, for example, there were
a significant population of higher density small objects with no
density measurements, the large objects could easily be made. Density 
measurement requires the presence of a satellite. It is not impossible
to imagine that perhaps less dense objects preferentially 
acquire satellites, but such a scenario seems contrived. Similarly,
2002 UX25 could be an outlier and not representative of the densities
of the mid-sized KBOs. More density measurements in this size range are
clearly warranted.

Finally, it is possible that objects of the dwarf planet size evolve to 
their high densities through the effects of giant impacts. Indeed, Haumea
is thought to have lost much of its icy mantle, clearly leading to
an increase in density \citep{2007Natur.446..294B}, but giant impact modeling has not
found a way to lose sufficient ice to affect the density enough to
explain more than a small amount of the higher densities of the dwarf planets
\citep{2009ApJ...691L.133S,2010ApJ...714.1789L}. 

None of these alternatives appears likely. We are left in the uncomfortable
state of having no satisfying mechanism to explain the formation of the 
icy dwarf planets. While objects up to the size of 2002 UX25 can 
easily be formed through standard coagulation scenarios, the rock rich larger
bodies may require a formation mechanism separate from the rest of the Kuiper belt.

 {\it Acknowledgments:}
This research has been supported by grant
NNX09AB49G from the NASA Planetary Astronomy program. 
Some of the data presented herein were obtained at the W.M. Keck Observatory, which is operated 
as a scientific partnership among the California Institute of Technology, the University of California and the 
NASA. The Observatory was made possible by the generous financial support of the W.M. Keck Foundation. 
Additional data were obtained from HST. Support for programs 10545 and 10860 were 
provided by NASA through a grant from the Space
Telescope Science Institute, which is operated by the Association of the Universities for Research in Astronomy, Inc., under NASA contract
NAS 5-26555. We thank the anonymous referee for thought provoking comments
which substantially improved the presentation of this manuscript.

\begin{table}
\begin{center}
\caption{Separation of 2002 UX25 and its satellite}
\begin{tabular}{lrrll}
\tableline\tableline
date &  RA offset & dec offset & telescope/ & relative brightness \\
(UT) & (mas) & (mas) & instrument& \% \\
2453609.15758 &    70.2$\pm$0.3 &    -146$\pm$1 & HST/HRC & 8.9$\pm$0.5 \\
2453939.30187 &     - & - & HST/HRC & -\\
2453939.98322 &    22$\pm$1 & -85$\pm$1 & HST/HRC & $6.7\pm0.2$ \\
2453944.04838 &   -23.0$\pm0.9$   & 74$\pm$2 & HST/HRC & 7.3$\pm$0.4 \\
2453947.42261 &  - & - & HST/HRC & -\\
2453958.34814 &    70.8$\pm$0.5  &  -143.4$\pm$0.3 & HST/HRC & 8.8$\pm$0.2 \\
2453965.20996 &    34.8$\pm$0.8  &  -105.7$\pm$0.9 & HST/HRC & 9.3$\pm$0.2 \\
2456268.78992 &       74$\pm$3 &        -133$\pm$3 & Keck/NIRC2 & 7.3$\pm$0.4 \\

\tableline
\end{tabular}
\end{center}
\end{table}

\begin{table}
\begin{center}
\caption{Orbital parameters\tablenotemark{a}}
\begin{tabular}{ll}

\tableline
semimajor axis & 4770$\pm$40 km \\
 inclination  & 275.5$\pm$0.3 deg \\
 period & 8.3094$\pm$0.0002 days \\
 eccentricity & 0.17$\pm$0.03\\
 argument of perihelion &254$\pm$1 deg \\
 longitude of ascending node& 23.3$\pm$0.3 deg\\
 time of pericenter passage &JD 2453976.94$\pm$0.03\\
mass  & $1.25 \pm 0.03 \times 10^{20}$ kg\\
heliocentric orbit-satellite orbit angle & 65 deg \\

\end{tabular}
\tablenotetext{a}{Relative to J2000 ecliptic}
\end{center}
\end{table}

\end{document}